\documentclass[a4paper,11pt]{article}%
\usepackage{graphicx} 
\usepackage[english]{babel}
\usepackage{authblk}
\usepackage{bbm}
\usepackage{caption}
\usepackage{tikz}
\usepackage{booktabs}
\usepackage{braket}
\usepackage{multirow}
\usepackage{soul}
\usetikzlibrary{trees}
\usetikzlibrary{decorations.pathmorphing}
\usetikzlibrary{decorations.markings}
\usepackage{tikz,ifthen}
\tikzset{beamerprimary/.style={structure.fg, thick}}
\tikzset{beamersecondary/.style={structure.bg, thick}}
\tikzset{boson/.style={draw=structure.fg,decorate, decoration={snake}},
    gauge/.style={decorate, decoration={snake} },
    fermion/.style={postaction={decorate},
        decoration={markings,mark=at position .55 with {\arrow{>}}}},
    fermionloop/.style={postaction={decorate},
        decoration={markings,mark=at position .25 with {\arrow{<}}}}, 
    gluon/.style={decorate, 
        decoration={coil,amplitude=4pt, segment length=5pt}},
    scalar/.style={dashed},
    scalarloop/.style={dashed={decorate},
        decoration={markings,mark=at position .25 with {\arrow{<}}}},
    resonance/.style={double,double distance=1.5pt}	
}
\usetikzlibrary{arrows,shapes,positioning}
\usetikzlibrary{decorations.markings}
\tikzstyle arrowstyle=[scale=1]
\tikzstyle directed=[postaction={decorate,decoration={markings,
    mark=at position .65 with {\arrow[arrowstyle]{stealth}}}}]
\tikzstyle reverse directed=[postaction={decorate,decoration={markings,
    mark=at position .65 with {\arrowreversed[arrowstyle]{stealth};}}}]

\usepackage[utf8]{inputenc}
\RequirePackage[T1]{fontenc}

\RequirePackage{flushend}
\RequirePackage{color}
\usepackage{changes}
\usepackage{multirow}
\usepackage{hyperref} 
\hypersetup{colorlinks,citecolor=blue,linkcolor=blue,urlcolor=blue}

\usepackage{ragged2e} 
\usepackage{subfigure} 
\usepackage{slashed}\usepackage{float}
\usepackage{amsmath,amssymb,lmodern,setspace}
\usepackage{multirow} 
\usepackage{cite}
\usepackage[capitalise]{cleveref}

\title{Axial-vector exchange contribution to the Hyperfine Splitting}

\author[1]{Alejandro Miranda~\footnote{jmiranda@fis.cinvestav.mx}}
\author[1]{Pablo Roig~\footnote{proig@fis.cinvestav.mx}}
\author[2]{Pablo Sánchez-Puertas~\footnote{psanchez@ifae.es}}

\affil[1]{\small Departamento de F\'isica, Centro de Investigaci\'on y de Estudios Avanzados del IPN,
Apdo. Postal 14-740,07000 Ciudad de M\'exico, M\'exico}
\affil[2]{Institut de F\'isica d'Altes Energies (IFAE),
The Barcelona Institute of Science and Technology (BIST),
Campus UAB, E-08193 Bellaterra (Barcelona), Spain}

\date{}

\begin{document}
\maketitle
\abstract{
We revisit the contribution of axial-vector mesons to the hyperfine splitting of muonic hydrogen. We focus our attention on the doubly-virtual asymptotic behavior of the relevant form factors of axial-vector mesons, together with their coupling to nucleons based on resonance saturation and short-distance constraints. Among others, we find significant differences with respect to previous studies, including an opposite sign and a $\sim50\%$ effect of the doubly-virtual high-energy behavior. }

\section*{Introduction}

The electromagnetic interactions of axial-vector mesons have attracted much attention recently. In particular, in the context of the hadronic light-by-light (HLbL) contribution to the anomalous magnetic moment of the muon~\cite{Pauk:2014rta, Jegerlehner:2017gek, Roig:2019reh, Leutgeb:2019gbz,Cappiello:2019hwh,Masjuan:2020jsf,Aoyama:2020ynm,Szczurek:2020hpc,Zanke:2021wiq,Colangelo:2021nkr,Leutgeb:2021mpu}, but also concerning  their contribution to the hyperfine structure (HFS) of muonic hydrogen~\cite{Dorokhov:2017nzk, Dorokhov:2019bwt}.

In the present study, we revise different aspects of their role in the HFS, briefly discussing axial-vector mesons decays into $
\ell^+\ell^-$ that enter the HFS calculation. 
On the one hand, we analyze the role of the high-energy behavior. This was missing in previous pioneering studies of the HFS~\cite{Dorokhov:2017nzk, Dorokhov:2019bwt}, but has been found to play an important role in the context of the HLbL~\cite{Masjuan:2020jsf,Leutgeb:2019gbz,Cappiello:2019hwh}. We find that the impact is by no means negligible, representing a $50\%$ effect. On the other hand, we use short-distance constraints connecting the Compton scattering tensor and the axial form factor of the nucleon. This allows to unambiguously fix the sign of the HFS contribution and to better understand potential off-shell effects~\cite{Dorokhov:2017nzk, Dorokhov:2019bwt}
. Overall, we obtain a value with opposite sign with respect to previous estimates that, together with the non-negligible impact of the high-energy behavior, represents the main result of this work. 
Besides, a discussion concerning the uncertainties on the relevant coupling constants and off-shell effects complements this paper. 

The article is organized as follows: in Sect. \ref{sec:Atoll}, we discuss the amplitude for $A\to\ell^+\ell^-$ decays, a necessary ingredient in our calculation. 
Building on the former, \cref{sec:HFS} outlines the contribution to the HFS on a general basis. The particular models are outlined in \cref{sec:model} based on resonance saturation. The final results and conclusions, including the impact on the Zemach radius are given in \cref{sec:ResCon}. Further information, including the form factor description, is relegated to the appendices.

\section{$A\to\ell^+\ell^-$ decays}\label{sec:Atoll}

The axial-vector meson decays to a lepton pair play a central role in computing their contribution to the HFS, to be discussed in the section below. Furthermore, they can provide important information regarding $A\to\gamma^*\gamma^*$ transitions~\cite{Rudenko:2017bel,Zanke:2021wiq}(see also the comments at the end of this section). 
We outline next the evaluation of the relevant matrix element appearing in these decays, which comparison to existing results will provide an additional (intermediate) cross-check of our evaluation.

The aforementioned  process occurs through the electromagnetic interactions and involves the $A\to\gamma^{*}\gamma^{*}$ transition, which can be expressed on the basis of 
Lorentz invariance and $C\!P$ symmetry as~\cite{Roig:2019reh}\footnote{We use $\epsilon^{0123}=+1$. The interested reader is referred to Ref.~\cite{Roig:2019reh} for relations to other bases. Comparing to the basis in \cite{Dorokhov:2017nzk}, $A_4 -\bar{A}_3 = B_2$, $\bar{A}_4 -{A}_3 = \bar{B}_2$, $2C_S = A_3 +\bar{A}_3$, $2C_A =A_3 -\bar{A}_3$, as well as $F_{AV\gamma^*\gamma^*}^{(0)}(q_1^2,q_2^2) = -B_{2S}(q_1^2,q_2^2) $. Also, comparing to the basis in \cite{Hoferichter:2020lap,Zanke:2021wiq}, $m_A^2 B_2 =-\mathcal{F}_3$, $m_A^2 \bar{B}_2 =\mathcal{F}_2$, $m_A^2 C_A =\mathcal{F}_1$. In addition, the form factors with well-defined symmetry are related by $2m_A^2 B_{2S}=\mathcal{F}_s$, $-2m_A^2 B_{2a}=\mathcal{F}_{a_2}$, $m_A^2 C_{A}=\mathcal{F}_{a_1}$.
}
\begin{multline}\label{eq:A2gamma}
i\,\mathcal{M}_{A\to\gamma^{*}\gamma^{*}}= i e^2\bigg\lbrace B_2(q_1^2,q_2^2)\,i\epsilon_{\mu\alpha\rho\beta}\,q_{1}^{\beta}\left[q_2^\alpha q_{2\nu}-g^\alpha_\nu\,q_2^2\right]+B_2(q_2^2,q_1^2)\,i\epsilon_{\nu\alpha\rho\beta}\,q_{2}^{\beta}\left[q_1^\alpha q_{1\mu}-g^\alpha_\mu\,q_1^2\right]\\
+i\epsilon_{\mu\nu\alpha\beta}\,q_1^\alpha q_2^\beta \left[\bar{q}_{12\rho} \,C_{A}(q_1^2,q_2^2) +q_{12\rho}\, C_{S}(q_1^2,q_2^2)\right] \bigg\rbrace\,\epsilon^{*\mu}(q_1) \epsilon^{*\nu}(q_2) \epsilon^{\rho}(q_{12})
\equiv ie^2\mathcal{M}_{A\,\mu\nu\rho} \epsilon^{*\mu}(q_1) \epsilon^{*\nu}(q_2) \epsilon^{\rho}(q_{12})
,
\end{multline}
where $q_{12}=q_{1}+q_{2}=q$ and $\bar{q}_{12}=q_{1}-q_{2}$. Here, $\epsilon^{*\mu}(q_1)$ and $\epsilon^{*\nu}(q_2)$ are the polarization vectors of the photons, while $\epsilon^\rho(q)$ is the polarization vector of the axial-vector meson with  $A=a_1,f_1^{(\prime)}$. 
Importantly, the basis in \cref{eq:A2gamma} is free of kinematic singularities, see also~\cite{Hoferichter:2020lap}. The form factors, $B_{2}(q_1^2,q_2^2)$, $C_{A}(q_1^2,q_2^2)$ and $C_{S}(q_1^2,q_2^2)$, encode the strong interaction dynamics. To guarantee Bose symmetry, $C_{A}(q_1^2,q_2^2)$ must be antisymmetric and $C_{S}(q_1^2,q_2^2)$ must be symmetric under $q_{1}\leftrightarrow q_{2}$. The contribution from $C_{S}$ vanishes when the axial-vector meson is on-shell and, in this basis, can be omitted when considering high-energy constraints~\cite{Masjuan:2020jsf}, which is not necessarily the case in other bases (see also Refs.~\cite{Hoferichter:2020lap,Zanke:2021wiq}). In the last expression, $C_{A}$ corresponds to transverse photons ($TT$) and $B_2$ is a combination of $TT$ and $LT$ polarization states ($L$ standing for longitudinal).

\begin{figure}[tbp]
\begin{center}
		\includegraphics[width=0.33\textwidth]{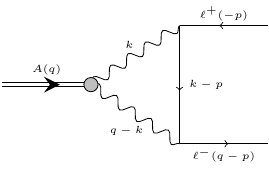}\hspace{1.5cm}
		\includegraphics[width=0.30\textwidth]{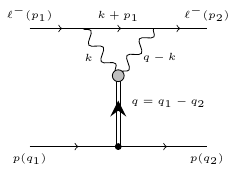}
    \captionsetup{width=0.88\linewidth}
\caption{\small The leading contribution to $A\to\ell^+\ell^-$ decays (left). The axial-vector meson contribution to the $\ell^- p \to \ell^- p$ amplitude relevant to the HFS (right). The grey blob includes structure-dependent axial-photon-photon interactions.}\label{fig1:All}
\end{center}
\end{figure}

The leading order contribution to $A\to\ell^+\ell^-$ decays is given by the diagram shown in fig.~\ref{fig1:All} (left), which corresponding amplitude can be expressed by means of Eq.~(\ref{eq:A2gamma}) as
\begin{equation}
    i\mathcal{M} = -e^4 \varepsilon_{\rho} \int \frac{d^4 k}{(2\pi)^4} \frac{\bar{u} \gamma_{\color{black}\nu}[(\slashed{k}  -\slashed{p})+m_{\ell}]\gamma_{\color{black}\mu}v}{q_1^2 q_2^2 [(k-p)^2 -m_{\ell}^2]}\mathcal{M}_A^{\mu\nu\rho}(q_1,q_2) \, ,
\end{equation}
with $q_1=k$ and $q_2=q-k$. In the following, it will be useful to express the most general amplitude for these decays, that based on Lorentz invariance and $C\!P$ symmetry can be written as
\begin{equation}\label{eq:A2LLdec}
i\mathcal{M}= i\,\bar{u}(q-p)\left[A_{1}(q^2)\,\gamma^\rho+A_{2}(q^2)\,q^\rho\right]\gamma^5 v(p)\,\epsilon_{\rho}(q) 
\equiv i\mathcal{M}^{\rho}_{A\to\bar{\ell}\ell}\epsilon_{\rho}(q).
\end{equation}
Note that the $A_2$ amplitude is a pure off-shell effect and, as such, it does not contribute to the decay width, while we keep it here as it will generally contribute to the Compton-scattering tensor that appears in the HFS. Using the projector techniques defined in \cref{sec:A1A2} and introducing 
$l=p_{\ell^-} -p_{\ell^+}$, we find for the $A_i(q^2)$  amplitudes
\begin{align}\small\label{eq:A1}
A_{1}(q^2)&=\frac{\alpha^2}{i\pi^2}\frac{1}{l^2\,q^2}\int d^4 k \frac{C_{A}(q_1^2,q_2^2)\,\omega_{A}+B_{2S}(q_1^2,q_2^2)\,\omega_{2S}+B_{2A}(q_1^2,q_2^2)\,\omega_{2A}}{q_1^2\,q_2^2[(k-p)^2-m_\ell^2]}, \\
\omega_{^{2S}_{2A}}&=\pm(q_1^2\pm q_2^2)\lbrace l^2(q\cdot q_1)(q\cdot q_2)-q^2(k\cdot l)[q_1\cdot q_2-q^2]\rbrace - l^2q^2\left\lbrace ^{2q_1^2q_2^2}_{0}\right\rbrace, \nonumber\\
\omega_{A}&=(q_1^2-q_2^2)\lbrace -q^2\,(k\cdot l)\,(q_1\cdot q_2) +l^2[k^2\,q^2-(k\cdot q)^2]\rbrace, \nonumber
\end{align}
together with
\begin{multline}\small\label{eq:A2}
A_{2}(q^2)=-\frac{2m_\ell}{q^2}A_{1}(q^2) + \frac{\alpha^2 }{i\pi^2}\frac{4m_\ell}{q^4}\int d^4 k \frac{k^2\,q^2-(k\cdot q)^2}{k^2(q-k)^2[(k-p)^2-m_\ell^2]}\left\lbrace - q^2\,C_S(q_1^2,q_2^2)\right.\\ 
\left.-(q_1^2-q_2^2) \left[C_A(q_1^2,q_2^2)-B_{2A}(q_1^2,q_2^2)\right]-(q_1^2+q_2^2)B_{2S}(q_1^2,q_2^2)\right\rbrace.
\end{multline}
In the previous equations, we have used form factors with well-defined symmetry following Refs.~\cite{Roig:2019reh,Masjuan:2020jsf}: 
$B_2(q_1^2,q_2^2)=B_{2S}(q_1^2,q_2^2)+B_{2A}(q_1^2,q_2^2)$ and $B_2(q_2^2,q_1^2)=B_{2S}(q_1^2,q_2^2)-B_{2A}(q_1^2,q_2^2)$.

Noteworthy, the current evaluation allows to cross-check our results for $A_1(m_A^2)$ against those in Refs.~\cite{Rudenko:2017bel,Zanke:2021wiq}, finding a nice agreement and reinforcing our results, to be used below in the $q^2\to0$ limit for the HFS. 

Finally, we would like to comment on an important  aspect. Namely, that $A\to e^+e^-$ decays are particularly sensitive to the intermediate $V\gamma$ contributions (and thereby to the timelike region), showing less sensitivity to high-energies or the spacelike regime. This is a consequence of the Landau-Yang theorem and is opposite to $\pi(\eta)\to \ell^+\ell^-$ decays \cite{Masjuan:2015cjl}, where the imaginary part is dominated by the intermediate $2\gamma$ state. Due to this reason, and the fact that several form factors appear (in contrast to the HFS where the knowledge of $B_{2S}$ suffices), we refrain from discussing this further. Still, we use different models for the $B_{2S}$ form factor (see \cref{sec:FFs}) to illustrate our claim for the $f_1(1285)$ case. 
In particular, taking the unpolarized spin-averaged squared matrix element $\overline{\mathcal{M}^2}$
and the corresponding partial decay width
\begin{equation}
\overline{\mathcal{M}^2}=\frac{4}{3}q^2\beta_\ell^2\vert A_{1}(q^2)\vert^2, \qquad
\Gamma_{A\to\ell\ell}=\frac{1}{12\pi}\,M_{A}\,\beta_{\ell}^3\vert A_{1}(M_A^2)\vert^2, \quad (\beta_\ell^2=1-4m_\ell^2/s) \, ,
\end{equation}
we find the results in \cref{tab:BR} using the form factors discussed in \cref{sec:FFs}.
\begin{table}[tbp]\centering\small
\begin{tabular}{ccccccc}\toprule
     & VMD & eVMD & heVMD & DIP & heDIP & OPE \\ \midrule
    $\mathcal{B}_{e^+e^-}$ & $1.90(^{92}_{74})$ 
    & $1.55(^{50}_{38})$  & $1.66(^{45}_{42})$  & $2.87(^{3.69}_{1.73})$  & $2.73(^{3.86}_{1.69})$  & $2.67(^{3.99}_{1.75})$ \\ \bottomrule
\end{tabular}
\captionsetup{width=0.88\linewidth}
\caption{
\small Branching fraction for $f_1\to e^+e^-$ decays in units of $10^{-9}$ with the different form factors outlined in \cref{sec:FFs} (ideal mixing case). In particular the first three columns correspond to models incorporating a vector meson mass $m_V=0.77$~GeV, whereas the last three columns have effective masses around $1$~GeV, illustrating the relevance of the intermediate $V\gamma$ state. For reference, this branching ratio is $<9.4\times10^{-9}$ at 90$\%$ confidence level~\cite{SND:2019rmq}.}
\label{tab:BR}
\end{table}
From the results therein, we find that the form factors including an explicit vector meson mass of $m_V=0.77$~GeV (VMD, eVMDm, heVMD) display similar results, with mild corrections from their different   high-energy behavior. On the contrary, they differ substantially from those employing an effective mass that successfully describes the L3 data~\cite{Achard:2001uu,Achard:2007hm} in the  (singly-virtual) spacelike region, regardless of their high-energy behavior. As we will show, this is the opposite for the HFS that, as such, 
might not benefit  substantially from a precise knowledge of $A\to e^+e^-$ decays.

\section{The contribution to the HFS}\label{sec:HFS}


Having computed the $\mathcal{M}^{\rho}_{A\to\bar{\ell}\ell}(q^2)$ amplitude in \cref{eq:A2LLdec}, the contribution of axial-vector mesons to the HFS is straightforward. In particular, the relevant amplitude of the $\ell^-p \to \ell^-p$ process driven by axial-vector mesons, \cref{fig1:All} (right), can be expressed as
\begin{align}\label{eq:HFSamp}
i\mathcal{M}_{\ell p} &= 
    ig_{ANN}[\bar{u}(A_1 \gamma^{\mu} +A_2 q^{\mu})\gamma^5 u]_{\ell} \frac{-g_{\mu\nu} +\frac{q_{\mu}q_{\nu}}{m_A^2}}{q^2 -m_A^2}  [\bar{u}\gamma^{\nu}\gamma^5 u]_N,
\end{align}
where we have introduced the coupling of the axial-vector mesons to the nucleons, $g_{ANN}$, via
\begin{equation}
\mathcal{L}_{a_1 NN} = -g_{a_1 NN} (\bar{N}\gamma_{\mu}\gamma^5 \vec{\sigma} N) \vec{a}_1^{\mu},    \quad
\mathcal{L}_{f_1 NN} = -g_{f_1 NN} (\bar{N}\gamma_{\mu}\gamma^5 N) f_1^{\mu}.
\end{equation}
Determining the couplings above will be an important part of our study, that we postpone to \cref{sec:model}.  
Pursuing further the nonrelativistic potential for the HFS, and making use of the relation $\mathcal{M}_{\ell p} = -2m_{\ell}2m_N \tilde{V}_{NR}(\boldsymbol{q}^2)$, we obtain\footnote{We use $\bar{u}(p_2,s_2) \gamma^5 u(p_1,s_1) \!\xrightarrow{\textrm{NR}}\! (\boldsymbol{p}_1 -\boldsymbol{p}_2)\!\cdot\! 
    [\xi_{s_2}^{\dagger}\boldsymbol{\sigma} \xi_{s_1}]$ and $\bar{u}(p_2,s_2) \gamma^{\mu}\gamma^5 u(p_1,s_1) \!\xrightarrow{\textrm{NR}}\!
   2m [\xi_{s_2}^{\dagger}\!\left(0,\boldsymbol{\sigma}\right) \xi_{s_1}]$, where $\boldsymbol{p}_1 -\boldsymbol{p}_2 \to \pm\boldsymbol{q}$ for nucleons(leptons). 
   }
\begin{multline}
    \tilde{V}_{NR}(\boldsymbol{q}^2) =  g_{ANN}\Bigg[ 
    \frac{A_1(-\boldsymbol{q}^2)}{m_A^2 +\boldsymbol{q}^2}\left\lbrace
    (\hat{\boldsymbol{\sigma}}_{\ell}\cdot
    \hat{\boldsymbol{\sigma}}_{N})
    +\frac{(\boldsymbol{q}\cdot\boldsymbol{\hat{\sigma}}_{\ell}) (\boldsymbol{q}\cdot\boldsymbol{\hat{\sigma}}_{N})}{m_A^2}
    \right\rbrace 
    -\frac{\tilde{A}_2(-\boldsymbol{q}^2)}{m_A^2}(\boldsymbol{q}\cdot\boldsymbol{\hat{\sigma}}_{\ell}) (\boldsymbol{q}\cdot\boldsymbol{\hat{\sigma}}_{N})
    \Bigg]\, , \label{eq:nonrelV}
\end{multline}
where $\hat{\boldsymbol{\sigma}}_{\ell(N)}$ are  Pauli matrices acting on the lepton(nucleon) spinors and $2m_\ell\tilde{A}_2=A_2$. 
In the following, we restrict ourselves to the leading-order contribution in $\alpha$. This justifies, in analogy with \cite{Dorokhov:2017nzk}, to neglect the terms proportional to $(\boldsymbol{q}\cdot\boldsymbol{\hat{\sigma}}_{\ell}) (\boldsymbol{q}\cdot\boldsymbol{\hat{\sigma}}_{N})$, as well as to take $A_1(-\boldsymbol{q}^2) \to A_1(0)$, both effects being suppressed by $m_{\ell}\alpha/\Lambda$ (see \cref{sec:HOV}). Furthermore, this justifies to keep with the leading term in the spinors' nonrelativistic expansion~\cite{Frugiuele:2021bic}.
Neglecting those terms, the expression above corresponds to a nonrelativistic potential
\begin{equation}
    \tilde{V}_{\textrm{NR}}(\boldsymbol{q}^2) \simeq  g_{ANN} 
    \frac{A_1(0)}{m_A^2 +\boldsymbol{q}^2}
    (\hat{\boldsymbol{\sigma}}_{\ell}\cdot
    \hat{\boldsymbol{\sigma}}_{N}), \qquad 
    V_{\textrm{NR}}(r) = \frac{g_{ANN}A_1(0)}{4\pi r}e^{-m_A r}
    (\hat{\boldsymbol{\sigma}}_{\ell}\cdot
    \hat{\boldsymbol{\sigma}}_{N})\,. \label{eq:NRpot}
\end{equation}
This agrees with the recent study in Ref.~\cite{Frugiuele:2021bic} upon identifying their coupling constants $g_{A}^{(1)}\to A_1(0)$, $g_A^{(2)} \to -g_{ANN}$. 
The corresponding shifts for each level can be obtained through 
$\Delta E = \bra{\Psi_{n,l,m}} V_{\textrm{NR}}(r) \ket{\Psi_{n,l,m}}$, with $\Psi_{n,l,m}$ the hydrogen atom wavefunctions. In particular, for the HFS we are interested in, corresponding to the energy difference $E(nS_{1/2}^{F=1}) - E(nS_{1/2}^{F=0})$~\cite{Frugiuele:2021bic}, it leads to:
\begin{align}
    \Delta E_1^{\textrm{HFS}} = 
    \frac{g_{ANN}A_1(0)}{\pi}
    \frac{(\mu\alpha)^3}{m_A^2}
    \frac{1}{(1+\frac{2\mu\alpha}{m_A})^2}
    \langle \hat{\boldsymbol{\sigma}}_{\ell}\cdot
    \hat{\boldsymbol{\sigma}}_{N} \rangle_{(\Delta F)} = 
    \frac{g_{ANN}A_1(0)}{\pi}
    \frac{(\mu\alpha)^3}{m_A^2}
    \frac{4}{(1+\frac{2\mu\alpha}{m_A})^2} \ , \label{eq:HFS1}\\ 
    \Delta E_2^{\textrm{HFS}} = 
    \frac{g_{ANN}A_1(0)}{16\pi}
    \frac{(\mu\alpha)^3}{m_A^2}
    \frac{2 +\big(\frac{\mu\alpha}{m_A}\big)^2}{(1+\frac{\mu\alpha}{m_A})^4}
    \langle \hat{\boldsymbol{\sigma}}_{\ell}\cdot
    \hat{\boldsymbol{\sigma}}_{N} \rangle_{(\Delta F)}
    = 
    \frac{g_{ANN}A_1(0)}{4\pi}
    \frac{(\mu\alpha)^3}{m_A^2}
    \frac{2 +\big(\frac{\mu\alpha}{m_A}\big)^2}{(1+\frac{\mu\alpha}{m_A})^4} \ ,  \label{eq:HFS2}
\end{align}
for $n=1,2$, where $\mu$ is the reduced mass, and the factor of $4$ in the right-hand side arises from the spin expectation value. We note that $A_1(0)$ can be expressed following the notation in Ref.~\cite{Dorokhov:2017nzk} as 
\begin{equation}\label{eq:A1Dorokhov}
    A_1(0) = \frac{4}{3}\left( \frac{\alpha}{\pi} \right)^2 \int_0^{\infty} dk^2 L_{\ell}(k^2) B_{2S}(-k^2,-k^2), 
\end{equation}
with $L_{\ell}(k^2)$ defined in Ref.~\cite{Dorokhov:2017nzk} (see Eq.~(14) therein).\footnote{We further note that, for the dipole (DIP) parametrization employed in Ref.~\cite{Dorokhov:2017nzk}, $A_1(0) = \frac{4}{3} \left( \frac{\alpha}{\pi} \right)^2  B_{2S}(0,0) I (m_{\ell})$, with $I({m_{\ell}})$ defined in the Eq.~(27) from Ref.~\cite{Dorokhov:2017nzk}.} The previous results show that only the $B_{2S}$ form factor contributes to the HFS to leading order in $\alpha$, simplifying the calculation as compared to $A\to e^+e^-$ decays. Likewise, it is straightforward to check that the general results in Ref.~\cite{Dorokhov:2017nzk} amount to our \cref{eq:HFS1,eq:HFS2} times a factor of $(-2)$. 
While we could not trace the factor of $2$, the relative sign appears comparing to their Eqs.~(5,20). Still, the sign depends on their photon momentum flow and $\epsilon^{0123}$ convention, that are unclear. More importantly, the final sign arising from \cref{eq:HFS1,eq:HFS2,eq:A1Dorokhov} will depend on the relative sign for $B_{2S}(0,0)$ and $g_{ANN}$, that was fixed in Ref.~\cite{Dorokhov:2017nzk} on the basis of quark-loop models. In the following section, we introduce our model to compute the HFS, that unambiguously fixes the sign in a transparent manner, finally confirming our opposite sign for the numerical results.
In any case, our agreement with Refs.~\cite{Rudenko:2017bel,Zanke:2021wiq} regarding $A_1(m_A^2)$, and Ref.~\cite{Frugiuele:2021bic} in deriving the nonrelativistic potential, further reinforces our findings, \cref{eq:HFS1,eq:HFS2}.


\section{
Model results
}\label{sec:model}

In order to obtain a numerical estimate for the HFS, determining the $g_{ANN}$ couplings is almost as important as fixing the sign of $B_{2S}(0,0)g_{ANN}$. In the following, we use short-distance constraints, that allow to relate the nucleon Compton scattering tensor to the nucleon axial form factors in a transparent manner. This allows to fix the sign and, eventually, obtain the desired couplings within a resonance saturation scheme. 
In particular, the relevant short-distance constraint follows from the operator product expansion (OPE) of two vector currents in the limit where $q_1^2\sim q_2^2\sim \hat{q}^2 
\gg \{ q_{12}^2, \Lambda_{\textrm{QCD}}^{2}\}$, where we have introduced $\hat{q} \equiv (q_1 -q_2)/2$ and $q_{12}=q_1 +q_2$. This reads~\cite{Melnikov:2003xd,Masjuan:2020jsf}:
\begin{equation}\label{eq:OPEcurrent}
    \int d^4x d^4y e^{i(q_1\cdot x +q_2\cdot y)}  T\{ j^{\mu}(x) j^{\nu}(y)\}   =  \frac{-2}{\hat{q}^2} \epsilon^{\mu\nu\alpha \hat{q}} \int d^4z e^{i q_{12}\cdot z}  j_{5\alpha}(z)  +\mathcal{O}\left(\frac{\Lambda_{\textrm{QCD}}^{2}}{\hat{q}^{2}}\right),
\end{equation}
with $j_5^{\mu} = \bar{q}\gamma^{\mu}\gamma^5 \mathcal{Q}^2 q$, $\epsilon^{\mu\nu\rho q_i}\equiv \epsilon^{\mu\nu\rho \alpha}q_{i\alpha}$, and $\epsilon^{0123}=1$. Note actually that, since the typical momentum in the atomic system is of $\mathcal{O}(m_{\ell}\alpha)$, this is indeed the relevant limit in this calculation when the loop momentum in \cref{fig1:All} (right) is large. Regarding the axial-vector meson form factor, this implies~\cite{Roig:2019reh,Masjuan:2020jsf,Hoferichter:2020lap,Zanke:2021wiq}
\begin{equation}\label{eq:B2SOPE}
    \lim_{\hat{q}^2\to\infty} \hat{q}^4B_{2S}(\hat{q}^2,\hat{q}^2) = \sum_a \operatorname{tr}(\mathcal{Q}^2\lambda^a)m_A F_A^a,
\end{equation}
where we have introduced the axial decay constant $\bra{0}\bar{q}\gamma^{\mu}\gamma^5\frac{\lambda^a}{2}q \ket{A} = F_A^a m_A$. This fixes $\operatorname{sgn}B_{2S}(0,0) = \operatorname{sgn}F_A m_A$ provided the form factor does not change sign in the spacelike region (which is the case here and in Ref.~\cite{Dorokhov:2017nzk}), thus reducing the problem to determine the sign for $F_A^a m_A g_{ANN}$. The latter combination appears indeed in the axial form factors of the proton ($a=3,8,0$),
\begin{equation}\label{eq:axffdef}
\big\langle p(k^\prime)\big\vert \bar{q}\gamma_\mu\gamma^5\lambda^{a} q\big\vert p(k)\big\rangle=\bar{u}(k^\prime)\left[\gamma_\mu G_{\mathcal{A}}^{a}(q^2)+\frac{q_\mu}{2m_N}G_{\mathcal{P}}^{a}(q^2)\right]\gamma^5 u(k)\, , 
\end{equation}
when adopting a resonance saturation scheme\label{eq:axffdef}. In particular, one finds~\cite{Masjuan:2012sk}
\begin{equation}\label{eq:ResSat}
    G_{\mathcal{A}}^a(q^2) = \sum_{A} \frac{2F_{A}^a m_{A} g_{ANN}}{m_{A}^2 -q^2},
\end{equation}
where the sum goes over the (infinite number of) axial-vector meson resonances. As we shall show, this ultimately allows to fix $\operatorname{sgn} g_{ANN} m_A F_A^a$ in terms of $G_A^a(0)$, which sign is well-known. Ultimately, the previous modelling guarantees to fulfill the corresponding OPE constraint for the Compton scattering tensor
\begin{equation}\label{eq:Compton2Axial}
\lim_{\hat{q}^2\gg \{q_{12}^2,\Lambda_{\textrm{QCD}}^2\}}
\int d^4x \ e^{iq_1\cdot x}
    \bra{p(k')} T\{j^{\mu}(x) j^{\nu}(0)\} \ket{p(k)} 
    = \frac{-2}{\hat{q}^2} \epsilon^{\mu\nu\alpha \hat{q}}
    \bra{p(k')}
    j_{5\alpha}(0)
    \ket{p(k)}
\end{equation}
provided \cref{eq:B2SOPE} is satisfied 
. In the following, we discuss the results obtained when truncating the sum in \cref{eq:ResSat} with either one or two resonances.

\subsection{One-resonance saturation}\label{secc:1res}

First, we start truncating the sum in \cref{eq:ResSat} with the lightest resonance. Then, the value of the coupling constants can be determined in terms of $G_A^a(0)$ as follows 
\begin{align}
    G_A^3(0) =g_A^3 &{}= 2g_{a_1NN}\frac{F_A}{m_{a_1}}\,, \label{eq:ga3}\\
    G_A^8(0) = \frac{g_A^8}{\sqrt{3}} &{}= 2F_A\Bigg[ \frac{g_{f_1NN}}{m_{f_1}}\cos(\phi-\phi_0) + \frac{g_{f'_1NN}}{m_{f'_1}}\sin(\phi-\phi_0)
    \Bigg], \label{eq:ga8}\\
    G_A^0(0) = \sqrt{\frac{2}{3}}g_A^0 &{}= 2F_A\Bigg[ -\frac{g_{f_1NN}}{m_{f_1}}\sin(\phi-\phi_0) + \frac{g_{f'_1NN}}{m_{f'_1}}\cos(\phi-\phi_0) \Bigg]\,, \label{eq:ga0}
\end{align}
with $\phi$ the $f_1-f_1'$ mixing angle in the flavor basis and $\phi_0=\arctan\sqrt{2}$ (cf. \cref{sec:FFs}).\footnote{In the basis from Refs.~\cite{Achard:2001uu,Achard:2007hm}, the relation is $\phi=\theta_A +\phi_0 -\pi/2$ which, for the mixing angle given there using $\gamma\gamma^*\to f^{(\prime)}_1$ reaction, results in $\phi=26.7(2)^{\circ}$. Recent studies~\cite{Du:2021zdg} suggest a range for the mixing angle $\phi\in(-7,23)^{\circ}$.} This implies (we adopt a  positive $F_A$), 
\begin{align}
    g_{a_1NN}&=5.6(1.1), &
    g_{f_1NN}&=2.01(0.17), &
    g_{f_1^{\prime}NN}&=-0.33(0.08), & (\phi&=0), \label{eq:gaIM}\\
    g_{a_1NN}&=5.6(1.1), &
    g_{f_1NN}&=1.93(0.16), &
    g_{f_1^{\prime}NN}&= 0.71(21), & (\phi_{\textrm{L3}}&=26.7(5.0)^{\circ}),
\end{align}
where we used $g_A^3=1.2730(13)$~\cite{Hayen:2021iga},  $g_A^8=0.530(18)$, $g_A^0=0.392(24)$~\cite{Alexandrou:2019brg}, $F_A=140(10)$~MeV~\cite{Dumm:2009va, Nugent:2013hxa, Masjuan:2020jsf} and the PDG~\cite{Zyla:2020zbs} masses with an additional uncertainty accounting for the half-width rule~\cite{Masjuan:2012gc}. The errors obtained for  $g_{a_1NN},\,g_{f_1NN},\,g_{f^{\prime}_1NN}$ are dominated by $m_{a_1}$, $F_A$, and $g_A^{8,0}$, respectively. 
Our results are similar to \cite{Dorokhov:2017nzk}, with a slight departure in the $f_1^{(\prime)}$ cases ---partly related to their use of the OZI rule (that in our scheme would require $g_A^8=g_A^0$). At this point, it is worth emphasizing that the ad hoc $1/e$ off-shell factor introduced in Ref.~\cite{Dorokhov:2017nzk} spoils the appropriate normalization for the axial form factors precisely at the $q^2\to0$ point and should be avoided. Further discussions on this point are 
included in the following section.

Having estimated the axial couplings, we move on to our results for $A_1(0)$.  Taking the models from \cref{sec:FFs}, we obtain the values in \cref{tab:A1pablo}. 
\begin{table}[tbp]\small
    \centering
    \begin{tabular}{ccccccc} \toprule
        & VMD & eVMD & DIP & heVMD & heDIP & OPE  \\ \midrule
   $f_1(1285)$ & $1.68(^{27}_{25})$  & $1.21(^{47}_{31})$ & $0.99(^{17}_{15})$ & $1.34(^{34}_{14})$ & $1.33(^{48}_{33})$ & $1.53(^{25}_{24})$ \\
   
   $a_1(1260)$ & $1.68(^{27}_{25})$  & $1.03(^{65}_{28})$ & $0.91(^{20}_{18})$ & $1.17(^{51}_{16})$ & $1.14(^{53}_{31})$ & $1.41(^{31}_{28})$ \\
   $f_1(1420)$ & $2.99(^{35}_{33})$  & $0.78(^{14}_{13})$ & $0.78(^{15}_{13})$ & $0.96(^{12}_{11})$ & $0.96(^{33}_{23})$ & $1.20(^{22}_{21})$ \\
   \bottomrule
    \end{tabular}
    \captionsetup{width=0.88\linewidth}
    \caption{
    \small The results for$A_1(0)/[\alpha^2 B_{2S}(0,0)]$ for $\ell=\mu$. For simplicity, we take ideal mixing in VMD models, implying that $m_V=0.77~\textrm{GeV} \simeq m_{\rho,\omega}$ for $a_1,f_1$ and $m_V=m_{\phi}$ for the $f_1'$.}
    \label{tab:A1pablo}
\end{table}
There, we find that models failing to incorporate the doubly-virtual high-energy $Q^2$ scaling (eVMD, DIP) underestimate the value for $A_1(0)$ ---even if correctly reproducing the singly-virtual L3 data. This is the case for the form factor in Ref.~\cite{Dorokhov:2017nzk}, that corresponds to our DIP column. This implies that in the present calculation one should employ only those form factors describing L3 data and incorporating the high-energy behavior (heVMD, heDIP, OPE). Among them, the OPE model represents our preferred choice since: (i) it reproduces L3 data~\cite{Achard:2001uu,Achard:2007hm}; (ii) it is the only one that fulfills the pQCD scaling for a large virtual photon regardless the second photon virtuality; (iii) for two virtual photons, it fulfills the OPE, \cref{eq:B2SOPE} (find further details in \cref{sec:FFs}). As such, we take it as the central value, incorporating the difference with respect to heVMD and heDIP models as an additional uncertainty.
Having determined the value for $A_1(0)$, we estimate the contribution of the lowest-lying axial-vector mesons to the HFS, that are collected in \cref{T7:HFSs}. 
\begin{table}[tbp]\small\color{black}
\begin{center}
\begin{tabular}{ccccc}
\toprule
\multirow{2}{*}{A} & \multirow{2}{*}{$\displaystyle\frac{A_1(0)}{\alpha^2 B_{2S}^{A}}$} 
& $B_{2S}^{A}(0,0)$ & $\Delta E^{HFS}_{A}(1S)$ & $\Delta E^{HFS}_{A}(2S)$ \\
 &  
 & $[\text{GeV}^{-2}]$ & $[\text{meV}]$ & $[\text{meV}]$ \\
\midrule
$f_1(1285)$ & $1.53(25)(^{+00}_{-20})$ 
& $0.269(30)$ & $0.011(2)(1)(1)(0)[0]$ & $0.0014(^{+2}_{-3})(1)(2)(0)[0]$ \\ 
$a_1(1260)$ &  $1.41(30)(^{+00}_{-27})$ 
& $0.245(63)$ &  $0.029(^{+6}_{-8})(6)(7)(2)[0]$ &  $0.0036(^{+8}_{-10})(7)(9)(2)[0]$ \\ 
$f_1(1420)$ & $1.20(22)(^{+00}_{-24})$ 
& $0.197(30)$ & $-0.001(0)(0)(0)(0)[^{+3}_{-0}]$ & $-0.0001(0)(0)(0)(0)[^{+3}_{-0}]$ \\ 
\midrule
TOTAL &  
&  &  $0.039(^{+12}_{-13})[^{+3}_{-0}]$ &  $0.0049(^{+14}_{-16})[^{+3}_{-0}]$ \\
\bottomrule
\end{tabular}
\captionsetup{width=0.88\linewidth}
\caption{\color{black}\small Results for the  HFS of muonic hydrogen. The central values for the $g_{ANN}$ couplings are those from ideal mixing, \cref{eq:gaIM}. The second column displays results from OPE column in \cref{tab:A1pablo}, including as an additional uncertainty the difference with other models therein (see details in the text). The final two columns include uncertainties from $A_1(0)$, $g_{ANN}$, $B_{2S}$, $m_A$ and an additional uncertainty from the mixing within brackets (see details in the text).
}
\label{T7:HFSs}
\end{center}
\end{table}
In the following section, we extend the model including an additional multiplet of axial-vector mesons. While this induces further model dependence concerning the transition form factors, it is known that at least two resonances are required to have a satisfactory description of the axial form factors of the nucleon. As such, it will serve as an estimate of our systematic uncertainties and to discuss off-shell effects.

\subsection{Two-resonance saturation\label{eq:secc2res}}

The one-resonance saturation employed in the previous section to describe the axial form factors of the nucleon and to estimate the $g_{ANN}$ couplings does not provide a satisfactory description of the axial form factor of the nucleon, that is better parametrized by a dipole form either in electroproduction~\cite{Bernard:2001rs} or lattice QCD data~\cite{Green:2017keo,Shintani:2018ozy,Jang:2019vkm,Alexandrou:2020okk,Bali:2019yiy,Park:2021ypf}. This can be partly understood on the basis of the high-energy behavior of the axial form factor, $\lim_{Q^2\to\infty}G_A^a(-Q^2)\sim \alpha^2_s(-Q^2)Q^{-4}$~\cite{Brodsky:1980sx,Carlson:1987en,Carlson:1987zr}, that requires the presence of at least two resonances to recover a  $Q^{-4}$ behavior~\cite{Masjuan:2012sk}. This suggests the necessity to go beyond  the one resonance saturation scheme, while this comes at the cost of non-negligible modeling 
of the poorly known heavy axial-vector meson resonances, including their masses and form factors. In order to estimate the masses of the heavier  multiplets, we use the Regge trajectory from Ref.~\cite{Masjuan:2012gc}: $m_{a_1(n)}^2 = m_{a_1}^2 + n\mu_3^2$, $m_{f_1^{(\prime)}(n)}^2 = m_{f_1^{(\prime)}}^2 + n\mu_0^2$, with $\mu_{3/0}^2= 1.36/1.19~\textrm{GeV}^2$. Imposing the normalization and the $Q^{-4}$ behavior of the axial form factors, we obtain the following coupling constants using ideal mixing
\begin{align}
    g_{a_1NN} &=11.8, &
    g_{f_1NN}&=4.78, &
    g_{f_1^{\prime}NN}&=-0.90, \\
    g_{a_1(1)NN} &=-8.6, &
    g_{f_1(1)NN}&=-3.64, &
    g_{f_1^{\prime}(1)NN}&=0.71.
\end{align}
The next part concerns the description of the $B_{2S}$ form factor of the heavy resonances. Lacking any experimental data, we resort to a Regge-like model from Ref.~\cite{Masjuan:2020jsf}
\begin{equation}\label{TFF:OPE2A}
B_{2S}^{A_n}(q_1^2,q_2^2)=\frac{B_{2S}^{A_n}(0,0)(M_a^2+n\Lambda^2)^2}{[q_1^2+q_2^2-(M_a^2+n\Lambda^2)]^2} \, ,\quad B_{2S}^{A_n}(0,0)= {{
\frac{B_{2S}^{A_0}(0,0)M_{a}^4 m_{A_n}}{(M_a^2+n\Lambda^2)^2 m_{A_0}} \, ,
}}
\end{equation}
that was created to describe some features of the $\langle VVA \rangle$ Green's function. As this induces further model dependence for the second multiplet ($n=1$), for which no data is available 
, we will use our results in this section to estimate systematic uncertainties in the one resonance 
saturation approach. Our results are given in \cref{tab:XX}.
\begin{table}[tbp]\small
\begin{center}
\begin{tabular}{cccccc}
\toprule
\multirow{2}{*}{A} & \multirow{2}{*}{$\displaystyle\frac{A_1(0)}{\alpha^2 B_{2S}^{A}}$} &  \multirow{2}{*}{$g_{ANN}$} & $B_{2S}^{A}(0,0)$ & $\Delta E^{HFS}_{A}(1S)$ & $\Delta E^{HFS}_{A}(2S)$ \\
 &  &  & $[\text{GeV}^{-2}]$ & $[\text{meV}]$ & $[\text{meV}]$ \\
\midrule
$f_1(1285)$ & $1.53$ & $4.78$ & $0.269$ & $0.0269$ & $0.0034$ \\
$f_{1}(\mathrm{1^{st}\;excitation})$ & $3.05$ & $-3.64$ & $0.093$ & $-0.0082$ & $-0.0010$ \\
Subtotal & & & & $0.0187$ & $0.0024$\\ \midrule

$a_1(1260)$ & $1.41$ & $11.8$ & $0.245$ & $0.0605$ & $0.0076$ \\ 
$a_{1}(\mathrm{1^{st}\;excitation})$ & $2.93$ & $-8.6$ & $0.082$ & $-0.0162$ & $-0.0020$ \\ 
Subtotal & & & & $0.0443$ & $0.0056$\\ \midrule

$f_1(1420)$ & $1.20$ & $-0.90$ & $0.197$ & $-0.0024$ & $-0.0003$ \\
$f_{1}^\prime(\mathrm{1^{st}\;excitation})$ & $2.72$ & $0.71$ & $0.051$ & $0.0007$ & $0.0001$ \\ 
Subtotal & & & & $-0.0017$ & $-0.0002$\\ \midrule

Total &  &  &  & $0.0613$ & $0.0078$ \\
\bottomrule
\end{tabular}
\captionsetup{width=0.88\linewidth}
\caption{\small
The contributions from the ground and first excited states contribution to the HFS (errors not included, see details in the text). The results compare to those in \cref{T7:HFSs}. The first resonance contribution is enhanced with respect to \cref{T7:HFSs} as a result of the $g_{ANN}$ coupling, whereas the first excited states partially damp this effect. 
}
\label{tab:XX}
\end{center}
\end{table}
We find that the enhanced couplings for the lowest-lying multiplet essentially double the HFS contribution with respect to the previous section. Such enhancement is partially cancelled by the contribution of the second multiplet, that reduces the final shift to a $60\%$ effect. Such variation could be taken as an off-shell effect, as it induces additional $q^2$ dependence besides the lowest-lying multiplet. However, its complexity goes beyond the $1/e$ factor in Ref.~\cite{Dorokhov:2017nzk} and a precise estimate would demand a better knowledge of the properties of the heavy axial-vector mesons, including their $g_{ANN}$ couplings and form factors. 

Given the large theoretical uncertainties in the results derived, especially owing to the  masses and form factors of the second multiplet, we stick to our results in the previous section and will assign the difference between the results in this and the previous subsection as an additional systematic uncertainty of our results. Overall, this points to a substantially larger contribution from the first multiplet and a partial reduction from heavier states.

\section{Results and conclusions}\label{sec:ResCon}

As our final result for the HFS, we take as our central value the result obtained with the one resonance saturation, incorporating as an additional systematic uncertainty the difference with respect to the two-resonance saturation approach. This gives
\begin{equation}\label{eq:HFSfinal}
    \Delta E^{HFS}_{A}(1S) = 0.039(^{+12}_{-13})(^{+3}_{-0})(^{+22}_{-00})~\textrm{meV}, \quad  \Delta E^{HFS}_{A}(2S) = 
    0.0049(^{+14}_{-16})(^{+3}_{-0})(^{+29}_{-00}) ~\textrm{meV}.
\end{equation}
Compared to Ref.~\cite{Dorokhov:2017nzk}, we find an opposite sign (and a factor of $2$ difference) in the calculation. Our results for the $A\to\ell^+\ell^-$ amplitude and the nonrelativistic expansion are in good agreement with existing studies, that further reinforces our findings. 
Besides, we find an important role (a $50\%$ effect roughly) of the doubly-virtual high-energy behavior of the transition form factor, that was one of our main goals in this study ---such effects should be included in future calculations of $\Delta E^{HFS}_{A}$.  

In addition, to fix the relevant signs of the form factors and coupling constants, we made use of the OPE. This provides a connection among the Compton scattering tensor and the axial form factors of the nucleon, that unambiguously defines the relevant signs when using a resonance saturation scheme. For the simplest scenario, that incorporates the lowest-lying resonance, we find similar couplings to those in Ref.~\cite{Dorokhov:2017nzk}, while substantial effects are found when two resonances are included. These are required to achieve a reasonable description of the axial form factors of the nucleon and points to a larger contribution of the lowest-lying multiplet together with a mild effect from the next one. The latter could be considered as an off-shell effect and discourages the use of ad hoc suppression factors as in \cite{Dorokhov:2017nzk}. The difference between the two scenarios is accounted for as an additional systematic uncertainty and points to the necessity of a better understanding of the nucleon to axial-vector meson couplings in order to improve in precision.  

\begin{figure}
    \centering
    \includegraphics[width=0.6\textwidth]{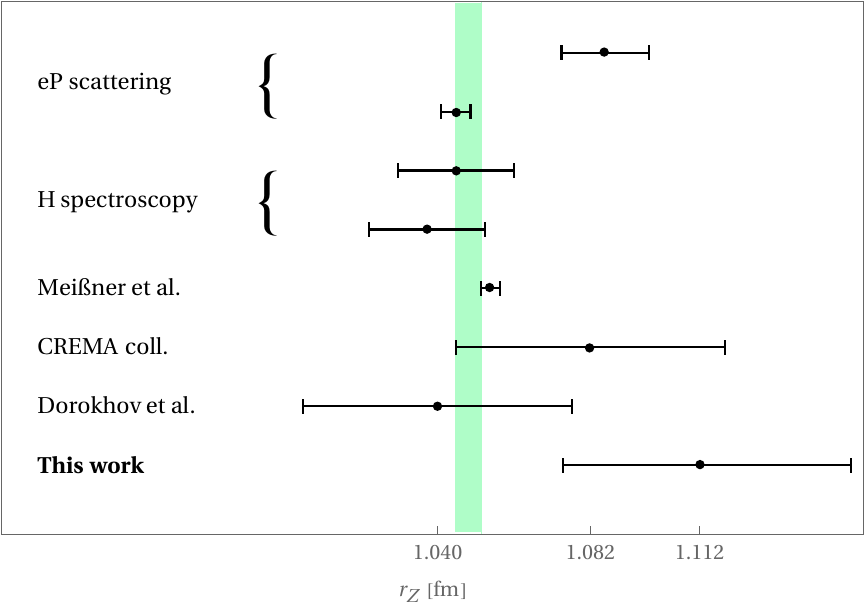}
    \captionsetup{width=0.88\linewidth}
    \caption{\small
    The Zemach radius ($r_Z$) from the references in the text and this work. The green band represents the average from Refs.~\cite{ Friar:2003zg,Distler:2010zq,Volotka:2004zu,Dupays:2003zz}.}
    \label{fig:Zemachr}
\end{figure}

Finally, we address the impact of this effect on the Zemach radius extraction by the CREMA Collaboration~\cite{Pohl:2010zza,Antognini:2013txn}, that measured the HFS of the $2S$ state, obtaining $\Delta E_{HFS}^{\mathrm{exp}}=22.8089(51)\,\text{meV}$. Comparing to the theoretical results for the HFS, $\Delta E_{HFS}^{\mathrm{th}}=22.9843(30)- 0.1621(10)\,r_{Z}\,\text{meV}$, see \cite{Faustov:2001pn,Martynenko:2004bt,Carlson:2011af,Borie:2011eia} and Table~3 from Ref.~\cite{Antognini:2013rsa}, they obtained $r_Z=1.082(37)\,\text{fm}$ \cite{Antognini:2013txn}. Incorporating the missing contributions from the axial vector mesons to the theoretical estimate in \cref{eq:HFSfinal} together with the pseudoscalar contribution \cite{Huong:2015naj}, $\Delta E^\pi_{HFS}=-(0.09\pm0.06)\,\mathrm{\mu eV}$, we obtain 
\begin{equation}
    r_Z= 1.112(31)_{\textrm{exp}}(19)_{\textrm{th}}(^{+20}_{-10})_{\textrm{axials}} \, . 
\end{equation}
The value is in mild tension with other estimates, $r_Z=1.086(12)\,\text{fm}$ \cite{Friar:2003zg}  and $r_Z=1.045(4)\,\text{fm}$ \cite{Distler:2010zq}, from electron-proton scattering, $r_Z=1.045(16)\,\text{fm}$ \cite{Volotka:2004zu}  and $r_Z=1.037(16)\,\text{fm}$ \cite{Dupays:2003zz}  from Hydrogen spectroscopy, and
$r_Z=1.054(3)\,\text{fm}$ \cite{Lin:2021xrc} 
 from electron-proton scattering and $e^+e^-$ annihilation.
We summarize these results in Fig. \ref{fig:Zemachr} where the green band corresponds to the average for electron-proton (eP) scattering and hydrogen (H) spectroscopy.

\section*{Acknowledgements}
P.~S.~P. thanks Clara Peset for interesting discussions on this topic. This work has been supported by the Spanish Ministry of Science and Innovation (grants SEV-2016-0588, FPA2017-86989-P and PID2020-112965GB-I00/AEI/10.13039/501100011033), by Secretaria d’Universitats i Recerca del Departament
d’Economia i Coneixement de la Generalitat de Catalunya under (grant 2017SGR1069), and European Union’s Horizon 2020 grants Research and Innovation Programme (grant no.~754510 (EU, H2020-MSCA-COFUND2016) and grant no.~824093 (H2020-INFRAIA-2018-1)). A. M. acknowledges Conacyt for his Ph.~D. scholarship and P.~R. the support of Cátedras Marcos Moshinsky (Fundación Marcos Moshinsky).

\appendix

\section{Projectors}\label{sec:A1A2}
The scalar functions $A_{1,2}(q^2)$ given in \cref{eq:A2LLdec} can be obtained by means of the following projection operators ($p_{1(2)}$ corresponds to the $\ell^-(\ell^+)$ momentum)
\begin{align}
    A_1(q^2)&{}=\frac{-1}{4(q^2-4m_\ell^2)}\mathrm{Tr}\left[\left(\slashed{p}_2-m_\ell\right)\left(\gamma_\rho+\frac{2m_\ell}{q^2}q_\rho\right)\gamma^5\left(\slashed{p}_1+m_\ell\right)\mathcal{M}^{\rho}_{A\to\bar{\ell}\ell}\right], \\
    A_2(q^2)&{}=\frac{m_\ell}{2 q^2(q^2-4m_\ell^2)}\mathrm{Tr}\left[\left(\slashed{p}_2-m_\ell\right)\left(\gamma_{\rho}-\frac{q^2-6m_\ell^2}{m_\ell\, q^2}q_{\rho}\right)\gamma^5\left(\slashed{p}_1+m_\ell\right)\mathcal{M}^{\rho}_{A\to\bar{\ell}\ell}\right] \\
    &{}=-\frac{2m_\ell}{q^2}A_1(q^2)-\frac{1}{2q^4}\mathrm{Tr}\left[\left(\slashed{p}_2-m_\ell\right) q_{\rho} \gamma^5\left(\slashed{p}_1+m_\ell\right)\mathcal{M}^{\rho}_{A\to\bar{\ell}\ell}\right].
\end{align}

\section{Form factors}\label{sec:FFs}


In this appendix, we describe the different models for the $B_{2S}(q_1^2,q_2^2)$ form factor used in the main text. Specifically, we discuss different variants in order to study the relevance of the asymptotic behavior. In particular, for the doubly-virtual symmetric kinematics one has the result in \cref{eq:B2SOPE} (see also Ref.~\cite{Hoferichter:2020lap}), enforcing $B_{2S}(-Q^2,-Q^2) \sim \mathcal{O}(Q^{-4})$ for large $Q^2$ values. In addition, in the singly-virtual kinematic regime, it is also known from the light-cone expansion that, for large $Q^2$ values, $B_{2S}(-Q^2,-q^2) \sim \mathcal{O}(Q^{-4})$, where $q^2 \ll Q^2$~\cite{Hoferichter:2020lap,Zanke:2021wiq}, that is also suggested by L3 data~\cite{Achard:2001uu,Achard:2007hm}.

The most simple form factor corresponds to the standard VMD prescription 
\begin{equation}
    B_{2S}^{\textrm{VMD}}(q_1^2,q_2^2) = \frac{B_{2S}(0,0) m_V^4}{(q_1^2 -m_V^2)(q_2^2 -m_V^2)},
\end{equation}
that, however, fails to describe the singly- and doubly-virtual asymptotic behavior, but is relevant to our discussion regarding $A\to e^+e^-$ decays. A variant that incorporates the appropriate high-energy behavior for singly-virtual kinematics is an extended VMD (eVMD) model with two resonances
\begin{equation}
    B_{2S}^{\textrm{eVMD/DIP}}(q_1^2,q_2^2) = \frac{B_{2S}(0,0) m_V^4 M^4}{(q_1^2 -m_V^2)(q_1^2 -M^2)(q_2^2 -m_V^2)(q_2^2 -M^2)},
\end{equation}
that still fails reproducing the OPE. A simplified variant of this model is the common dipole parametrization used in \cite{Achard:2001uu,Achard:2007hm,Dorokhov:2017nzk}, where $m_V=M$ and that we denote as DIP. We can amend this in a VMD incorporating the high-energy behavior (heVMD/heDIP) as follows
\begin{equation}
    B_{2S}^{\textrm{he(VMD/DIP)}}(q_1^2,q_2^2) = \frac{B_{2S}(0,0) m_V^4 M^4\left[ 1+ q_1^2 q_2^2 \Lambda_{OPE}^{-4} \right]}{(q_1^2 -m_V^2)(q_1^2 -M^2)(q_2^2 -m_V^2)(q_2^2 -M^2)}.
\end{equation}
Still, we note that such a form factor does not fulfill the appropriate high-energy behavior for $B_{2S}(-Q^2,-q^2)$ unless $q^2=0$. To better reproduce the high-energy behavior, we introduce the following form factor from Ref.~\cite{Masjuan:2020jsf} inspired in \cite{Masjuan:2017tvw}, that we label as OPE,
\begin{equation}
    B_{2S}^{\textrm{OPE}}(q_1^2,q_2^2) = \frac{B_{2S}(0,0)\Lambda_A^4}{(q_1^2 +q_2^2 -\Lambda_A^2)^2}.
\end{equation}
It describes L3 Collaboration results provided $\Lambda_A$ is chosen according to the dipole parameters in L3~\cite{Achard:2001uu,Achard:2007hm} and its doubly-virtual space-like behavior is in good agreement with the holographic results in Ref.~\cite{Leutgeb:2019gbz}, representing our preferred choice.

For the normalization, we take the values for $f_1,f_1^{\prime}$ from L3~\cite{Achard:2001uu,Achard:2007hm} together with our estimate in \cite{Roig:2019reh,Masjuan:2020jsf} for the $a_1$: $B_{2S}(0,0)=\{ 0.269(30), 0.197(30), 0.245(63) \}\textrm{GeV}^{-2}$ for $\{ f_1,f_1^{\prime},a_1 \}$. Regarding the mass parameter, we take both, for the OPE and (he)DIP variants, $m_V=M=\Lambda_A = \{1.04(8),0.926(79),1.0(1) \}$~GeV~, see Refs.~\cite{Achard:2001uu,Achard:2007hm,Roig:2019reh,Masjuan:2020jsf}. Concerning the eVMD and heVD models, we fix the $M$ parameter to reproduce the slope from the L3 Collaboration dipole in order to share the same low-energy behavior, which is accomplished adopting $M^2=\frac{\Lambda_A^2\,m_V^2}{2m_V^2-\Lambda_A^2}\sim 2$~GeV for $m_V=0.77$~GeV. Finally, to ensure the OPE behavior in \cref{eq:B2SOPE} in he(VMD/DIP) models, we find for ideal/L3 mixing
\begin{equation}
\Lambda_{OPE}^{f_1,f_1^\prime,a_1}/m_V M =\lbrace1.28(4)/1.37(5) ,\, 1.58(7)/1.26(6) ,\, 1.44(10)\rbrace\,\text{GeV}^{-1}, 
\end{equation}
respectively. In the equation above, we have employed the following mixing scheme
\begin{equation}
\left(\begin{array}{c}
f_1\\
f_1^\prime\\
\end{array}\right)=\left(\begin{array}{cc}
\cos\theta & -\sin\theta\\
\sin\theta & \cos\theta\\
\end{array}\right)\left(\begin{array}{c}
f^8\\
f^0\\
\end{array}\right),
\end{equation}
where $\theta$ is the mixing angle between the $SU(3)$ singlet ($f^0$) and octet ($f^8$) states.
Also, it is possible to write the last expression as
\begin{equation}
\left(\begin{array}{c}
f_1\\
f_1^\prime\\
\end{array}\right)=\left(\begin{array}{cc}
\cos\phi & -\sin\phi\\
\sin\phi & \cos\phi\\
\end{array}\right)\left(\begin{array}{c}
f^{NS}\\
f^{S}\\
\end{array}\right),
\end{equation}
where $\phi$ is the mixing angle between the non-strange ($f^{NS}$) and strange ($f^S$) states. $\theta$ and $\phi$ are related through $\theta=\phi-\phi_0$ with  $\phi_0=\arctan{\sqrt{2}}$ 
and the ideal mixing angle corresponds to $\phi=0$. The angles above relate to the one used in L3 Coll.~\cite{Achard:2001uu,Achard:2007hm} $(\theta_A=62(5)^\circ)$ as  
$\theta=\theta_A-\frac{\pi}{2}$ ($\phi=\theta_A+\phi_0-\frac{\pi}{2}$). 
In this study, and following Ref.~\cite{Roig:2019reh}, we take as our preferred value $\phi=0$, while we will take into consideration the L3 mixing angle as an additional uncertainty. Note also recent discussions concerning the mixing angle in Refs.~\cite{Du:2021zdg,Shastry:2021asu}. 

\section{Higher order effects in the nonrelativistic potential\label{sec:HOV}}

In this section we justify the suppression of the terms that have been neglected in evaluating the nonrelativistic potential in \cref{eq:nonrelV}. In particular, we start noticing the suppression corresponding to the potential of the kind $\tilde{V}_{NR}(\boldsymbol{q}^2) =
   (\boldsymbol{q}\cdot\boldsymbol{\hat{\sigma}}_{\ell}) (\boldsymbol{q}\cdot\boldsymbol{\hat{\sigma}}_{N})[m_A^2(m_A^2 +\boldsymbol{q}^2)]^{-1}$, 
   that in position space reads
\begin{align} 
V_{NR}(r) &=
   \frac{1}{3}\frac{\delta^{(3)}(r)}{m_A^2} 
    \langle \hat{\boldsymbol{\sigma}}_{\ell}\cdot
    \hat{\boldsymbol{\sigma}}_{N} \rangle
    -\frac{1}{3}\frac{e^{-m_A r}}{4\pi r}\left[ 
    S_{12}\left( 1 +\frac{3}{rm_A} +\frac{3}{(rm_A)^2} \right) +
        \langle \hat{\boldsymbol{\sigma}}_{\ell}\cdot
    \hat{\boldsymbol{\sigma}}_{N} \rangle
    \right], \nonumber \\
    &\Rightarrow \frac{1}{3}\left[ 
    \frac{\delta^{(3)}(r)}{m_A^2} -\frac{e^{-m_A r}}{4\pi r}
    \right]\langle \hat{\boldsymbol{\sigma}}_{\ell}\cdot
    \hat{\boldsymbol{\sigma}}_{N} \rangle, \label{eq:VNRqspin}
\end{align}
where in the last line we have ommitted $S_{12} = (3\hat{r}^i\hat{r}^j -\delta^{ij}) \hat{\boldsymbol{\sigma}}_{\ell}^i \hat{\boldsymbol{\sigma}}_{N}^j$, that is a rank-2 symmetric tensor and does not contribute to $S$-wave states. Accounting for this, the result reduces to the combination of the $\delta^{(3)}(r)$ contribution and the Yukawa part in \cref{eq:HFS1,eq:HFS2}. Noting that $|\Psi_{1(2),0,0}(0)|^2 = (\mu\alpha)^3/[(8)\pi]$, the cancellation of the Yukawa and $\delta$ terms in \cref{eq:VNRqspin} to leading order in $(\mu\alpha/m_A)$ is clear, with the final result reading
\begin{align}
\Delta E_1^{\textrm{HFS}} &=    \left[ \frac{4(\mu\alpha)^4}{3\pi m_A^3}\frac{1+\epsilon}{(1+2\epsilon)^2}\right]
   \langle \hat{\boldsymbol{\sigma}}_{\ell}\cdot
    \hat{\boldsymbol{\sigma}}_{N} \rangle_{(F=1-F=0)}, \\
\Delta E_2^{\textrm{HFS}} &=    \left[\frac{(\mu\alpha)^4}{48\pi m_A^3}\frac{8+ 11\epsilon +8\epsilon^2 +2\epsilon^3}{(1+\epsilon)^4}\right] 
   \langle \hat{\boldsymbol{\sigma}}_{\ell}\cdot
    \hat{\boldsymbol{\sigma}}_{N} \rangle_{(F=1-F=0)}, 
\end{align}
where $\epsilon=\mu\alpha/m_A$. With these results at hand, it is straightforward to show the suppression from the $A_1(\boldsymbol{q}^2)$ dependence. Noting $A_1(q^2)= A_1(0)  + \frac{q^2}{\pi}\int d\xi \frac{\operatorname{Im} A(\xi)}{\xi -q^2}$, the first term corresponds to our main result, whereas the second one leads to a potential of the kind $V(r) = \frac{1}{\pi}\int d\xi \operatorname{Im} A(\xi) \xi \left[ \frac{e^{-\sqrt{\xi}r}}{
4\pi
r} -\frac{\delta^{(3)}(r)}{\xi}  \right] $ that, in parallel with \cref{eq:VNRqspin}, is $\alpha$ suppressed. Note in addition that the lower threshold in the previous integral corresponds to the intermediate $V\gamma$ state, so one expects the relevant scale to be above $m_V$.

\bibliographystyle{hunsrt}
\bibliography{bib}
\addcontentsline{toc}{section}{References}
\end{document}